\documentclass[journal]{IEEEtran}
\usepackage{blindtext, graphicx}
\usepackage{cite}

\usepackage[cmex10]{amsmath}

\usepackage{url}

\usepackage{multirow}
\usepackage{color}

\usepackage[font = footnotesize]{caption}
\usepackage{graphicx}
\usepackage{subfig}
\usepackage{epstopdf}

\usepackage[table,svgnames]{xcolor} 

\usepackage{booktabs}

\begin{document}

\title{Transparent Tx and Rx Waveform Processing for \\
5G New Radio Mobile Communications}

\author{\IEEEauthorblockN{Toni Levanen\IEEEauthorrefmark{1}, Juho Pirskanen\IEEEauthorrefmark{2}, Kari Pajukoski\IEEEauthorrefmark{3}, Markku Renfors\IEEEauthorrefmark{1}, Mikko Valkama\IEEEauthorrefmark{1}}\\
\IEEEauthorblockA{\IEEEauthorrefmark{1}Laboratory of Electronics and Communications Engineering, Tampere University of Technology, Finland}\\
\IEEEauthorblockA{\IEEEauthorrefmark{2}Wirepas Oy, Finland}\\
\IEEEauthorblockA{\IEEEauthorrefmark{3}Nokia Bell Labs, Finland}}

\maketitle

\begin{abstract}

Several different waveform processing techniques have been studied and proposed for the 5G new radio (NR) physical layer, to support new mixed numerology and asynchronous services. The evaluation and comparison of these different techniques is commonly based on matched waveform processing in the transmitter and receiver units. In this article, it is shown that different techniques can be flexibly mixed, allowing to separately optimize complexity-performance trade-offs for transmitter and receiver implementations. Mixing of different waveform processing techniques is possible by setting adequate requirements for transmitter and receiver baseband processing allowing transparent design. The basic framework of how transmitter and receiver units can be independently applied and evaluated in the context of transparent design and an extensive set of examples of the achievable radio link performance with unmatched transmitter and receiver waveform processing are provided. The discussed approach and solutions simplify standardization, improve transparent transmitter and receiver coexistence, and allow future-proof evolution path for performance improvements in 5G NR physical layer hardware and algorithm design.

\end{abstract}

\begin{IEEEkeywords}
5G, New Radio (NR), PHY layer, radio link performance, receiver, system evolution, time-to-market, transmitter, transparent waveform processing, complexity
\end{IEEEkeywords}

\section{Introduction}
{\let\thefootnote\relax\footnotetext{\footnotesize This work has been accepted for publication in IEEE Wireless Communications Magazine. This is the revised version of the original work and it is currently in press. Copyright may be transferred without notice, after which this version may no longer be accessible.}}

The debate over different waveform processing solutions for the fifth generation (5G) new radio (NR) \cite{3GPPTS38300} physical layer air interface has been extensive during the last few years. Researchers have revised their knowledge over different waveforms thoroughly, evaluating, e.g., orthogonal and non-orthogonal multicarrier waveforms \cite{20145GNOW}, complex and real symbol spaces for subcarrier wise filtered multicarrier \cite{2014BanelliModFormatsAndWaveformsFor5G}, and cyclic prefix (CP) versus zero postfix \cite{2016_Coexistence_UFOFDM_CPOFDM} or zero tail combined with discrete Fourier transform spread orthogonal frequency division multiplexing (OFDM) \cite{2014_OFDM_enhancements_for_5g}. However, less studies have considered situations where transmitter (Tx) and receiver (Rx) units use different waveform processing techniques, which is the main theme of this article. 

The recent 3GPP technical report describing 5G NR \cite{3GPPTR38802}, states that the baseline assumption of the waveform for below 52.6~GHz communications is CP-OFDM and that the Tx processing has to be transparent to the Rx. This implies that any \textit{additional signal processing} on top of the commonly agreed baseline CP-OFDM waveform, e.g., time domain windowing or bandwidth part filtering performed in the Tx is not signaled to Rx and is thus generally unknown. Similarly, Rx waveform processing adopted on top of the basic CP-OFDM receiver processing is generally unknown to the Tx. In this article, we refer to Tx and Rx units which are capable of operating under such conditions as agnostic devices. Thus, it follows that Tx and Rx waveform processing stages are typically not matched, and that Tx and Rx waveform processing solutions need to be evaluated separately. By mixed transparent Tx and Rx waveform processing, we refer to scenarios where Tx and Rx use unmatched waveform processing techniques without information of the specific processing solution used by the other end. A concrete uplink (UL) example of such mixed or unmatched Tx and Rx waveform processing is a scenario where the user equipment (UE) uses time domain windowed overlap-and-add (WOLA) processing (see \cite{J:2013_Bala_WOLATxRx}, more details in Section \ref{sec:unit_testing_results}) in Tx and the next generation NodeB (gNB) uses channel filtering in the Rx. Several other example scenarios are also considered in this article, in both UL and downlink (DL), in the 5G NR system context.

In 5G NR networks, the concept of transparent waveform processing is especially important due to the new requirements to support inband mixed numerology as well as asynchronous traffic \cite{3GPPTR38802}, which do not exist in LTE. In the Tx side, the target of additional signal processing is to suppress the out-of-band emissions and inband interference leakage to achieve defined emission masks, adjacent-channel-leakage-ratio (ACLR) specifications, and good spectral efficiency when mixing different numerologies or asynchronous traffic. In the Rx side, the additional processing is used to improve adjacent-channel-selectivity which reduces the interference from a nearby intra-carrier or inter-carrier interferer using different transmission link direction, numerology, timing or completely different waveform. The mixed numerology inband emission masks considered for UL are a new aspect for 5G NR to allow in-channel mixing of different services using different numerologies and thus possibly different waveform processing within a carrier \cite{3GPPTR38803}. 
In general, the so-called \textit{bandwidth part} (BWP) is defined \cite{3GPPTS38300,3GPPTS38213} as a contiguous spectral block used to support certain radio interface numerology either in DL or UL, and is the main tool to provide support for mixed numerology or asynchronous UL services in 5G NR networks. By a \textit{subband}, in turn, we refer to a frequency allocation partially or fully covering one BWP. This reflects a minimum resource unit over which the windowing or filtering can be applied in the evaluated waveform processing methods. Such waveform processing seeks to specifically suppress the interference between different subbands or BWPs inside the carrier, arising from the adoption of multiple subcarrier spacings or relaxed time-synchronization requirements \cite{20145GNOW,J:2017YliKaakinenFC-F-OFDM}. 

In general, when the Tx and Rx units use different waveform processing techniques to shape the signal spectrum, special care in the design and evaluation of the used techniques is required. A particularly important aspect in applying transparent Tx waveform processing is that in \textit{interference-free scenarios}, the transmitted signal can still be received by a \textit{plain CP-OFDM Rx}, which refers to an Rx solution without any additional windowing or filtering on top of the baseline CP-OFDM waveform. The same holds for a transparent Rx, which has to work well with a plain CP-OFDM Tx. Ensuring a good performance with plain CP-OFDM Tx or Rx unit implies that different more evolved Tx and Rx units relying on transparent waveform processing can be flexibly taken into use in the evolving networks, over time, without any backward compatibility related concerns.

The observation that different Tx and Rx waveform processing algorithms can be flexibly mixed as long as they work efficiently also with plain CP-OFDM Rx or Tx, respectively, and that they provide similar performance with respect to matched Tx-Rx links is novel and generally less discussed in the available literature. Increasing awareness of these aspects is one of the main purposes of this article. In addition, the 5G NR radio link performance results and computational complexity comparisons presented in this article are novel and have not been presented earlier in such a variety. Furthermore, a common assumption in mixed numerology or asynchronous traffic evaluations has been that the desired and interfering signals are built using the same waveform processing techniques. This simplifying assumption becomes invalid when different spectrum enhancement techniques are gradually deployed in different 5G NR network entities, thus hindering the performance comparison of different waveform processing techniques. In this article, we describe a generalized performance evaluation framework that allows for a smooth long-term evolution of 5G NR devices without affecting the physical layer specification requirements, as long as the Tx and Rx units' waveform related transparent signal processing achieves the given selectivity characteristics and inband distortion requirements. Transparent waveform processing assumption also provides larger freedom in device design, such that different performance vs. complexity tradeoffs can be pursued. These are all aspects discussed in this article, increasing the novelty and technical impact of the work. 

The rest of the article is organized as follows. In Section \ref{sec:discussion}, the benefits of agnostic Tx and Rx framework and shifting from matched waveform processing oriented view to mixed processing based radio links in the context of 5G system evolution are shortly discussed. In Section \ref{sec:framework}, a generic evaluation framework for transparent Tx and Rx units based on the assumption of CP-OFDM based waveforms is described. Then, in Section \ref{sec:unit_testing_results}, the system parameterization is given followed by numerous performance results including several examples of mixed, transparent Tx and Rx processing cases in different channel environments and interference scenarios. In addition, the waveform processing related complexity evaluations and comparisons are provided in Section \ref{sec:unit_testing_results}. Finally, conclusions are drawn in Section \ref{sec:conclusions}.

\begin{figure} 
  \centering
  \includegraphics[angle=0,width=0.75\columnwidth]{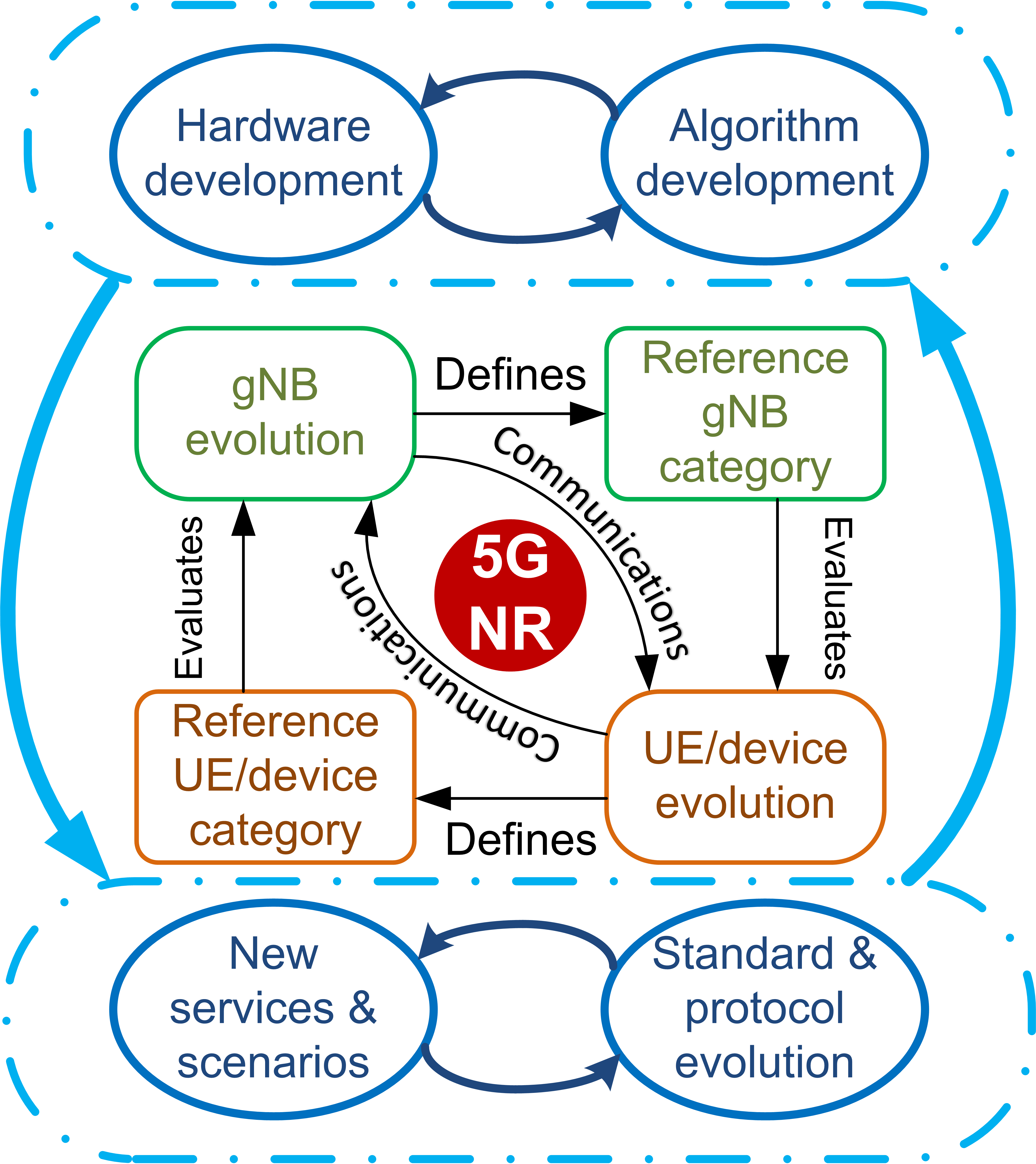}
  \caption{High level evolution circle of 5G NR.}
  \label{fig:evolutionCircle}
\end{figure}

\begin{figure*}
  \centering
  \subfloat[][]{\includegraphics[angle=0,width=0.95\columnwidth]{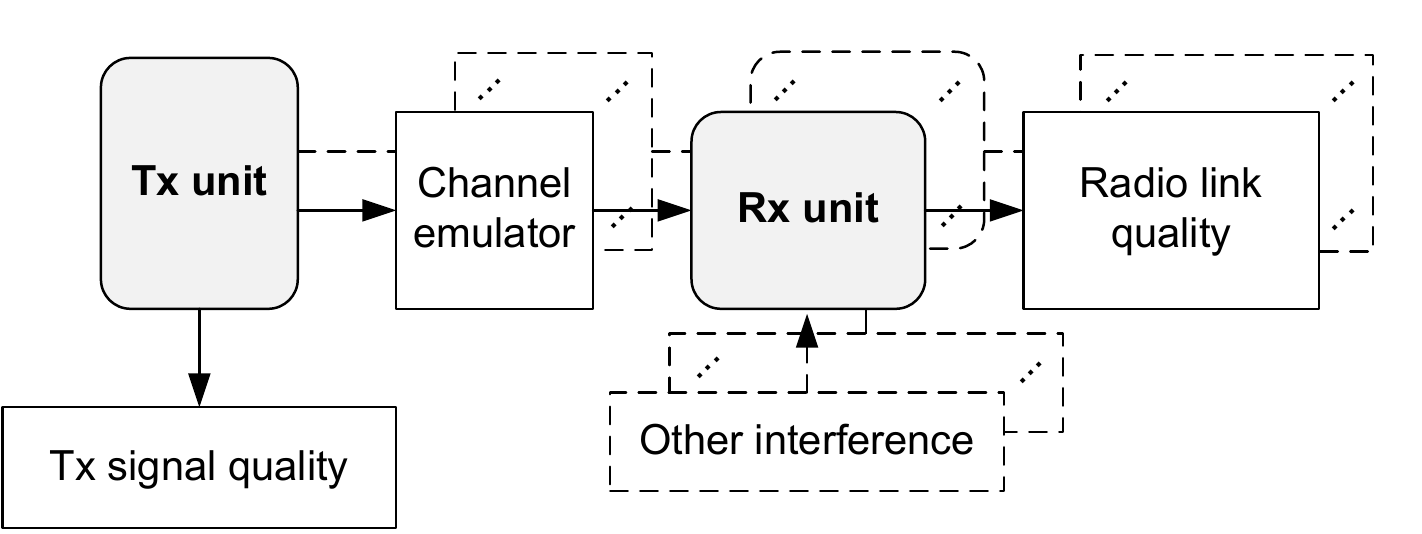}}
  \qquad
  \subfloat[][]{\includegraphics[angle=0,width=0.96\columnwidth]{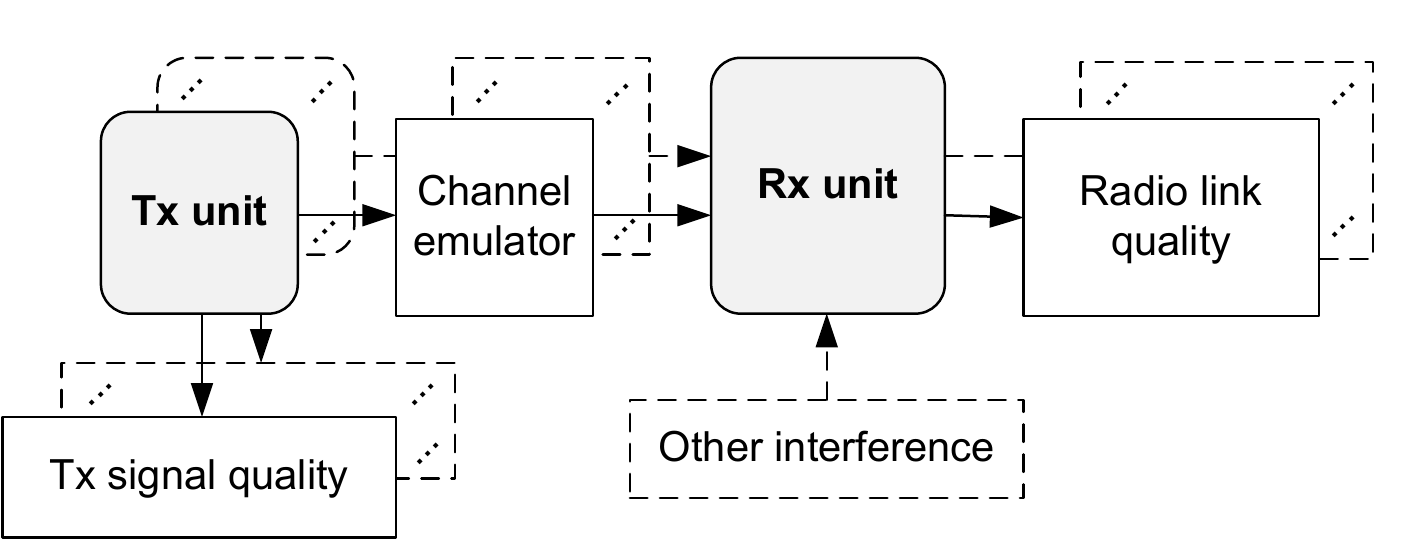}}
  \qquad
  \subfloat[][]{\includegraphics[angle=0,width=1.85\columnwidth]{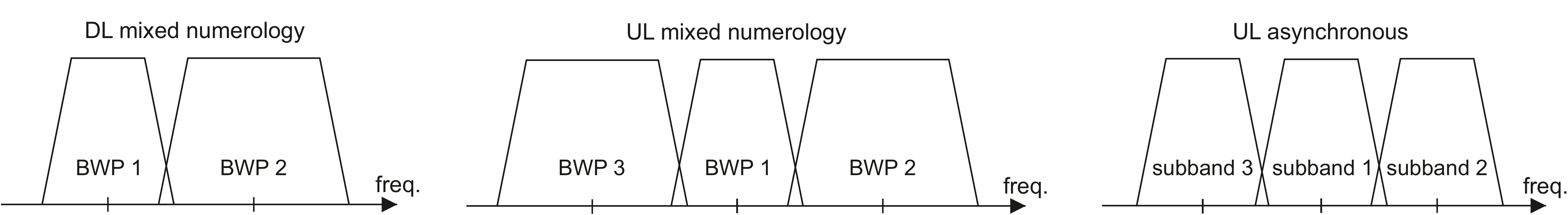}}
  \caption{The considered evaluation framework for different Tx and Rx units in (a) downlink and (b) uplink. In (c), the considered evaluation scenarios are shown with more detailed evaluation assumptions being listed in Table I. In this article, the radio link performance is always evaluated for BWP 1 or subband 1, depending on the scenario.}
  \label{fig:Tx_Rx_unit_testing}
\end{figure*}

\section{Supporting Flexible 5G NR System Evolution Through Transparent Tx and Rx Processing}
\label{sec:discussion}

The high level vision of 5G NR evolution in terms of commercial devices and reference Tx and Rx categories is illustrated in Fig. \ref{fig:evolutionCircle}. The gNBs and UEs or other devices (e.g. relays) communicate in the actual 5G NR networks. The evolution of gNB defines new reference gNB Tx and Rx categories that are used to test and evaluate UEs or other devices. In a similar manner, evolution of UEs or other devices defines new reference UE or device Tx and Rx categories which are used in evaluating gNBs. 5G NR evolution is, in general, driven by four main elements. Hardware development defines the improvements in hardware components and also improvements in computational capabilities of different network elements, allowing to use more sophisticated algorithms. Algorithm development, in turn, includes all new innovations in signal processing at different layers of communications. New services and scenarios refer to new markets of the future where the 5G NR physical layer is applied to unforeseen services. Standard and protocol evolution refers to standardizing new services and features which the forthcoming 5G NR releases have to support. Standardization work follows the evolution and predictions of new service markets and tries to adapt for those as quickly as possible. Also, standardization bodies are responsible for striving to improve the performance in existing services. Together, standard evolution and new revenue-increasing services drive the hardware and algorithm development. Then again, developments in hardware and algorithm domains drive the evolution of standards and enable new services. 

The agnostic Tx and Rx approach has several indirect future technology benefits. In standardization, hard exclusive decisions of the supported processing techniques can be avoided, allowing different chip-set manufacturers and device vendors to select different processing schemes in their implementation variants. Additionally, this approach allows fast time-to-market for the first 5G product designs on selected frequency bands where operators accept none or very limited amount of additional inband and out-of-band filtering compared to current LTE solutions. Even if additional frequency isolation is needed, this can be achieved simply by introducing additional guard bands by not using certain OFDM sub-carriers or physical resource blocks (PRBs) in scheduling, e.g., at the band edges or in BWPs using services with different physical layer numerology. Naturally this reduces system spectrum utilization efficiency but when the implementation techniques are improved, additional waveform signal processing can be applied separately at the network and the UE sides in fully backward compatible manner, \textit{without negative impacts on the existing UE population on the used band}. 

Eventually, it is expected that the extremely high traffic capacity requirements of the future, together with the need of supporting multiple services and dynamic time division duplexing will drive towards chip-set and device implementations supporting advanced signal processing techniques for suppressing inband and out-of-band interference and emissions in Rx and Tx sides, respectively. However, with the agnostic Tx and Rx approach, there is no pre-requisite to do so before launching 5G operation, and even more importantly there is no upper limit on the performance set by the standard. Instead, the upper limit is set by the cost versus benefits evaluation in real system deployments, and implementations can push this limit forward in the future years of 5G networks.  

\section{Agnostic Tx and Rx Unit Evaluation Framework}
\label{sec:framework}

\subsection{Reference Tx and Rx Assumptions}

The reference Tx and Rx designs assumed in this article reflect our view on the first device implementations in 5G NR phase 1, and are stemming from the current status of dominating LTE implementations. More specifically, we assume time domain WOLA processing \cite{J:2013_Bala_WOLATxRx} as the reference UE Tx solution. For UE Rx, no additional processing on top of plain CP-OFDM receiver is assumed. For the gNB Tx and Rx processing, channel filtered CP-OFDM is considered as the baseline. Typical LTE base station implementations rely on some type of channel filtering due to more strict DL Tx ACLR and out-of-band emission masks, and UL Rx adjacent-channel-selectivity requirements. Thus, in DL scenario the reference Tx is channel filtered CP-OFDM and the reference Rx is a plain CP-OFDM, while in UL the reference Tx is WOLA processed CP-OFDM and the reference Rx is a channel filtered CP-OFDM. All other more involved Tx-Rx processing combinations then reflect further evolution of the 5G NR networks and device categories. Note that in the selection of these reference Tx and Rx unit waveform processing techniques, we have extensively verified that the corresponding radio link performance with plain CP-OFDM Rx and Tx, respectively, is not degraded. In addition, the assumed reference Tx-Rx pairs in DL and UL are also already agnostic and unmatched solutions, by definition.

\subsection{Tx and Rx Unit Evaluation}

A block diagram of the Tx and Rx unit evaluation is shown in Fig. \ref{fig:Tx_Rx_unit_testing} (a) for DL and in (b) for UL. In Fig. \ref{fig:Tx_Rx_unit_testing} (a), the Tx unit corresponds to a gNB and the Rx unit(s) model UE Rx(s), and vice versa in (b). In Section \ref{sec:unit_testing_results}, we will evaluate several different radio link scenarios based on these models, with the above defined reference Tx and Rx solutions as the basis. The agnostic Tx and Rx solutions evaluated in Section \ref{sec:unit_testing_results}, correspond to the scenarios where we vary the waveform signal processing in the Tx or Rx unit, while we keep the reference waveform processing in the Rx or Tx unit, respectively. We also provide, for reference, selected matched link performance results in which the Tx and Rx units utilize the same waveform processing approach.

If the Tx units are capable of generating multiple BWPs with different subcarrier spacings, each of these BWPs is either assumed to be received by an independent Rx unit or all BWPs are received by a single Rx unit. This feature is new for 5G NR compared to LTE, in which only a single subcarrier spacing is used per transmitted signal. In the recent Rel-15 5G NR specification \cite{3GPPTS38300}, UEs are not mandated to transmit or receive multiple different numerologies simultaneously, but this requirement will most likely be included in later releases. In general, when multiple BWPs with different subcarrier spacings are transmitted from a single or more Tx unit(s), sufficient guard bands (GBs) need to be used to limit the interference between BWPs. The required GB and its minimization is one of the key drivers for introducing and adopting new waveform processing solutions in 5G NR Tx and Rx units.

In the baseline inband interference free evaluations, only a single Tx and Rx unit is considered without any inter-subband-interference. In the more evolved mixed numerology interference scenarios, in turn, we assume that several UEs are served by different BWPs. An example is illustrated in Fig. \ref{fig:Tx_Rx_unit_testing} (c), depicting the transmission of two (DL) or three (UL) BWPs as defined in \cite{3GPPTR38802}. In Fig. \ref{fig:Tx_Rx_unit_testing} (c), also the UL asynchronous interference scenario is illustrated, where within one BWP we have three distinct subbands that are assumed to be used by three asynchronous UEs. The main difference between the DL and UL Tx unit evaluations in different inband interference scenarios is that in DL the Tx unit applies the same waveform processing over all BWPs and they are combined in the baseband before the radio channel, whereas in UL each Tx unit can use any waveform processing solution and the transmitted signals are effectively combined in the gNB Rx antenna after the radio channels. In our UL evaluations, we vary only the waveform processing solution of a single Tx unit/BWP/subband of interest (BWP 1 or subband 1 in Fig. \ref{fig:Tx_Rx_unit_testing} (c)) while assume the reference Tx waveform processing in the other interfering Tx units/BWPs/subbands.

In general, the most important metric used to evaluate the quality of the received signal and the achievable throughput is the radio link block error rate (BLER), which is measured by the \textit{Radio link quality} block in Fig. \ref{fig:Tx_Rx_unit_testing}. The \textit{Tx signal quality} block, in turn, is used to evaluate the Tx error vector magnitude (EVM) and the spectral containment of the signal generated in the Tx unit. The spectral containment is measured in terms of out-of-band and inband emission levels as well as ACLR, which relates the absolute emissions in certain measurement band to the average power on the active band used by the Tx unit. In addition, the 5G NR Tx unit performance evaluation contains a new edge PRB based definition for Tx EVM and in the future possibly also mixed numerology specific inband emissions requirements \cite{3GPPTR38803}, which do not exist in current LTE standards. For realistic radio link performance evaluations, a \textit{Channel emulator} block is needed to introduce different standardized channel profiles, noise, frequency and power offsets, and timing errors. For generality, an additional element in the radio link performance evaluations is the \textit{Other interference} block, which can introduce additional standardized inband and out-of-band interference patterns under which the Rx unit has to achieve the specified performance metrics.

The described Tx and Rx unit performance evaluation setup is easily extended to evaluate carrier aggregation and multipoint transmission schemes, or even different baseline waveform types, e.g., discrete Fourier transform spread OFDM with CP or zero prefix.

\begin{table}[!t]
\renewcommand{\arraystretch}{1.2}
    \caption{Considered 5G NR physical layer parameterization \cite{3GPPTR38802} and assumed reference Tx and Rx solutions} 
    \label{tab:PHYParams}
    \centering
    \begin{tabular}{|l|c|}
        \hline
        Parameter & Value \\
        \hline
         Carrier frequency & 4~GHz \\
         Channel bandwidth & 10~MHz \\
         Sampling rate & 15.36~MHz\\
         FFT size ($N_{\text{FFT}}$) & 1024 \\
         CP length ($N_{\text{CP}}$) & 72 \\
         Subcarrier spacing & 15~kHz, 30~kHz \\
         \multirow{2}{*}{Channel model } & TDL-C 300~ns (DL, UL) \\
         & TDL-C 1000~ns (UL) \\
         UE mobility & 3 km/h \\
         \multirow{2}{*}{Modulation} & 256-QAM (DL, UL) \\
         & 64-QAM (UL)  \\
         Channel code & turbo code \\
         \multirow{2}{*}{Coding Rate} & 4/5 (DL, UL) \\
         & 3/4 (UL)  \\
         BLER target & 10\% \\
         Number of subcarriers per PRB & 12 \\
         Allocation granularity & 4~PRBs \\
         \hline
         \multicolumn{2}{|c|}{gNB Reference Tx and Rx Solutions} \\
         \hline
         Tx Waveform processing & channel filtered CP-OFDM\\
         Rx Waveform processing & channel filtered CP-OFDM\\
         Filter length & 165\\
         Stopband attenuation & 50 dB\\
         \hline
         \multicolumn{2}{|c|}{UE Reference Tx and Rx Solutions} \\
         \hline
         Tx Waveform processing & WOLA-based CP-OFDM\\
         Rx Waveform processing & plain CP-OFDM\\
         Window slope length & $N_{\text{CP}}/8=9$ \\
         \hline
         \multicolumn{2}{|c|}{WOLA} \\
         \hline
         Window slope length & $N_{\text{CP}}/8=9$ \\
         \hline
         \multicolumn{2}{|c|}{f-OFDM} \\
         \hline
         Filter type & Hann-windowed sinc-pulse \\
         Filter length & 512 \\
         Tone offset & 2 FFT bins \\
         \hline
         \multicolumn{2}{|c|}{FC-F-OFDM} \\
         \hline
         Transition bandwidth & 2 FFT bins \\
         Stopband minimum attenuation & 10 dB\\
         \hline
    \end{tabular}
\end{table}

\section{Example Unmatched Tx-Rx Scenarios, Performance Results and Complexity Aspects}
\label{sec:unit_testing_results}

In this section, the considered system parameterization is given, example performance results for different combinations of Tx and Rx units are provided and evaluated in terms of the achievable radio link BLER, and in the end a complexity comparison between evaluated waveform processing solutions is provided. The considered waveform processing solutions are fast convolution based subband filtered CP-OFDM (FC-F-OFDM) \cite{J:2017YliKaakinenFC-F-OFDM}, filtered CP-OFDM (f-OFDM) \cite{2015_Zhang_f-OFDM_for_5G}, and windowed overlap-and-add (WOLA) processed CP-OFDM \cite{J:2013_Bala_WOLATxRx}. In all cases, the assumed reference Tx-Rx radio link performance for DL or UL is also given as a baseline.

\subsection{Evaluation System Parameterization}
\label{subsec:system_parameterization}

The baseline physical layer definition and numerology follow the ones defined in \cite{3GPPTR38802} for 5G NR link level evaluations, and are summarized in Table \ref{tab:PHYParams}. The achievable Tx and Rx unit performance is evaluated in TDL-C channels \cite{3GPPTR38900} with 300~ns and 1000~ns root mean squared delay spreads. In addition, radio link performance with mixed numerology interference in DL and UL and with asynchronous interference in UL is evaluated. The mixed numerology and asynchronous interference scenarios follow the specified test-cases defined by 3GPP in \cite{3GPPTR38802} and are illustrated in Fig. \ref{fig:Tx_Rx_unit_testing} (c). The radio link performance is always evaluated for BWP 1 or sub-band 1 while other signals serve as inband interference sources. All signals are assumed to have 4 PRB allocation, and no other interference sources are assumed. Hence the word \textit{interference} refers below to the inband interference between different BWPs or subbands.

 FC-F-OFDM \cite{J:2017YliKaakinenFC-F-OFDM}, is an efficient and flexible subband or BWP filtering scheme that allows computationally efficient implementation of steep subband, BWP or channel filters. The filter design is based on optimized frequency domain windows allowing to balance the required minimum stopband attenuation, transition bandwidth, and EVM performance. These designs are characterized by the minimum stopband attenuation (typically at the stopband edge) and transition bandwidth \cite{J:2017YliKaakinenFC-F-OFDM}, which are selected here as 10~dB and 2~FFT bins (30~kHz), respectively.

 f-OFDM \cite{2015_Zhang_f-OFDM_for_5G} is based on time-domain subband or BWP filters which are built from windowed sinc-pulses with width in frequency corresponding to the subband or BWP size. The used time domain window is a Hann-window and the used filter length is 512 samples. For f-OFDM, tone offset defines the extra passband width with respect to the allocated bandwidth and is given in multiples of the subcarrier spacing.
 Tone offset is used to reduce the edge PRB EVM in f-OFDM. 

In the case of WOLA \cite{J:2013_Bala_WOLATxRx}, window slope length of 9 samples is used which corresponds to approximately 1\% rolloff. WOLA is a well known, low complexity technique to reduce out-of-band emissions of a CP-OFDM signal. In 5G NR, the new aspect of WOLA is to use it also in the Rx to reduce the mixed numerology or asynchronous interference effect on Rx demodulation quality. The Rx window is located in such a manner that the last window sample aligns with the last sample of the CP-OFDM symbol. This way the effective CP length is maximized with WOLA Rx processing. 

All the presented results assume an ideal channel knowledge in the Rx and each simulated slot contains only data symbols. A constant CP length is assumed for simplicity. For the DL, a modified Rapp power amplifier (PA) model \cite{2016RAN1RappPA} is used while for UL a polynomial PA model of order nine \cite{2016RAN1PolynomialPA} is adopted. These particular PA models are selected because they are used also by 3GPP and are publicly available. 

\subsection{Performance Evaluation Results}
\label{subsec:performance_evaluation_results}

\begin{figure*}
  \centering
  \subfloat[][]{\includegraphics[angle=0,width=0.95\columnwidth]{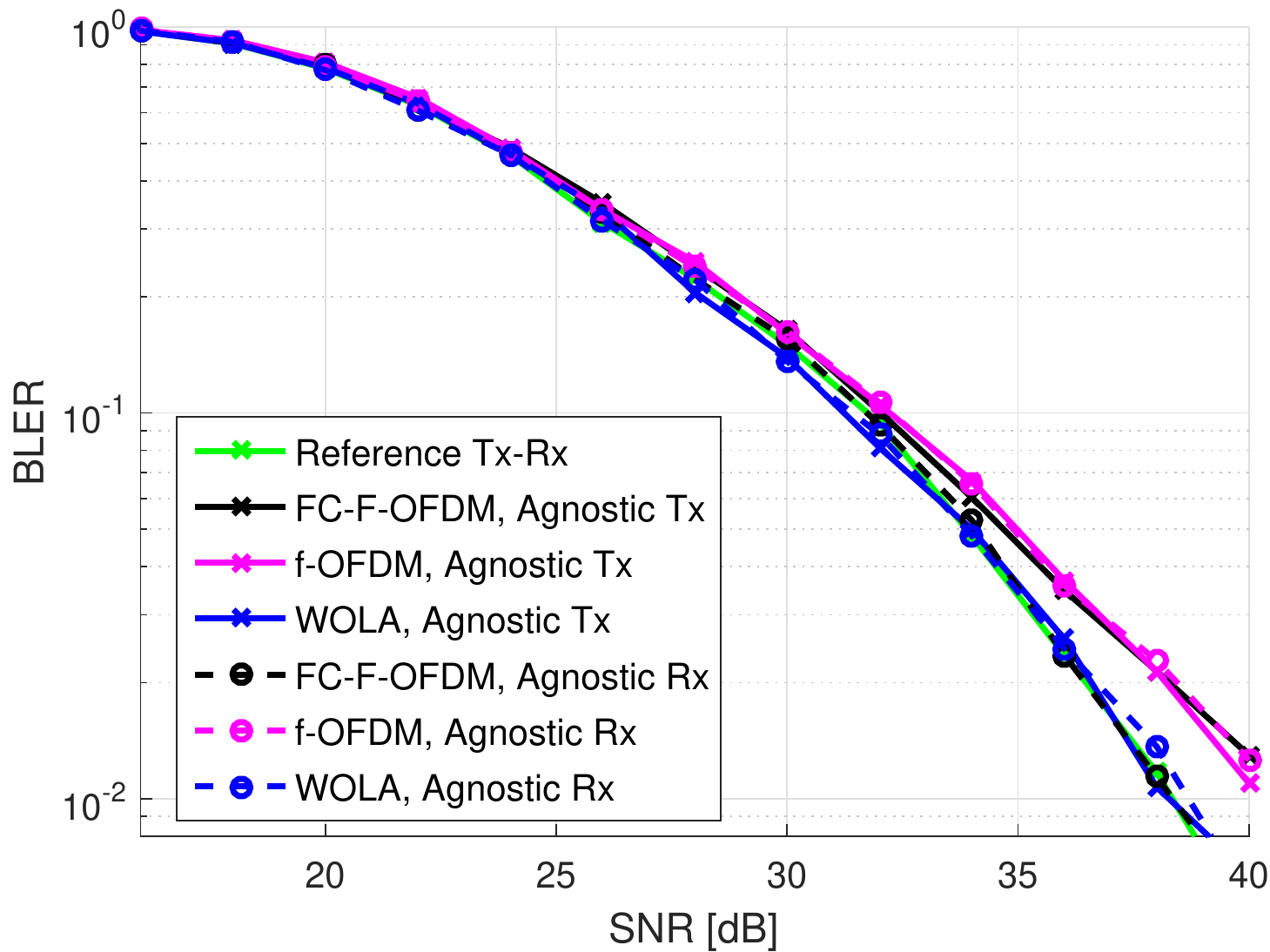}}
  \qquad
  \subfloat[][]{\includegraphics[angle=0,width=0.95\columnwidth]{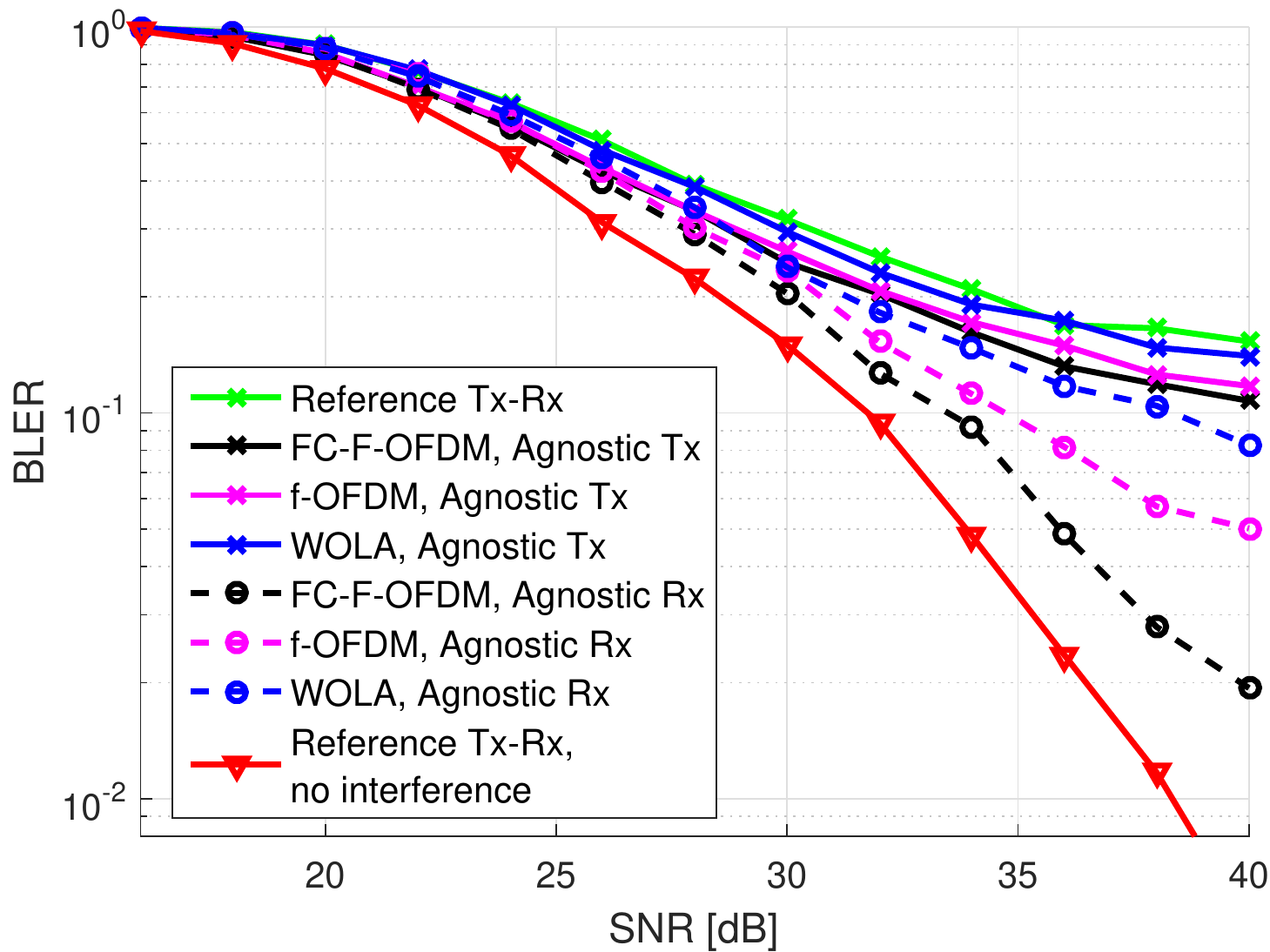}}
  \qquad
  \subfloat[][]{\includegraphics[angle=0,width=0.95\columnwidth]{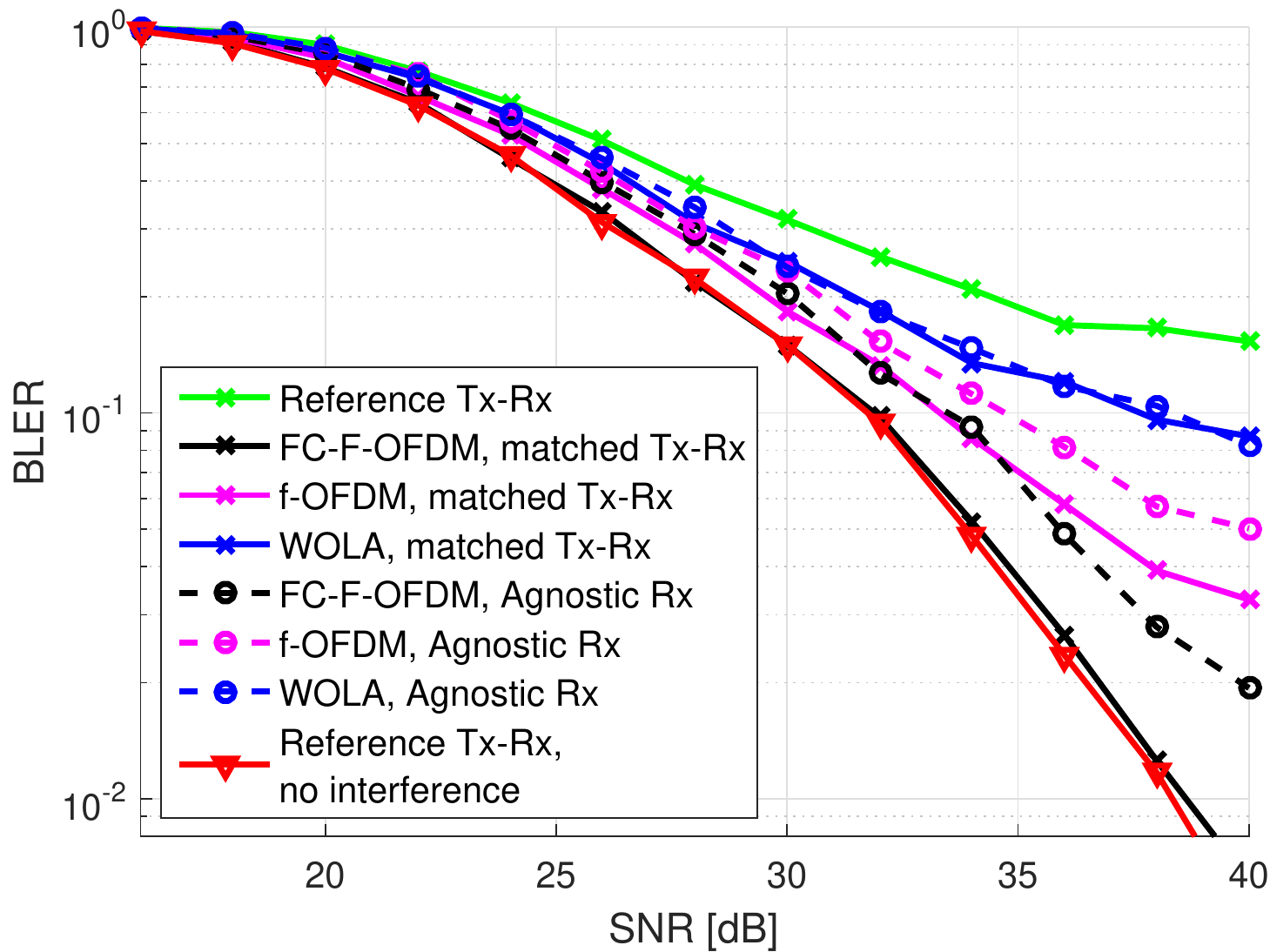}}
  \qquad
  \subfloat[][]{\includegraphics[width=0.95\columnwidth]{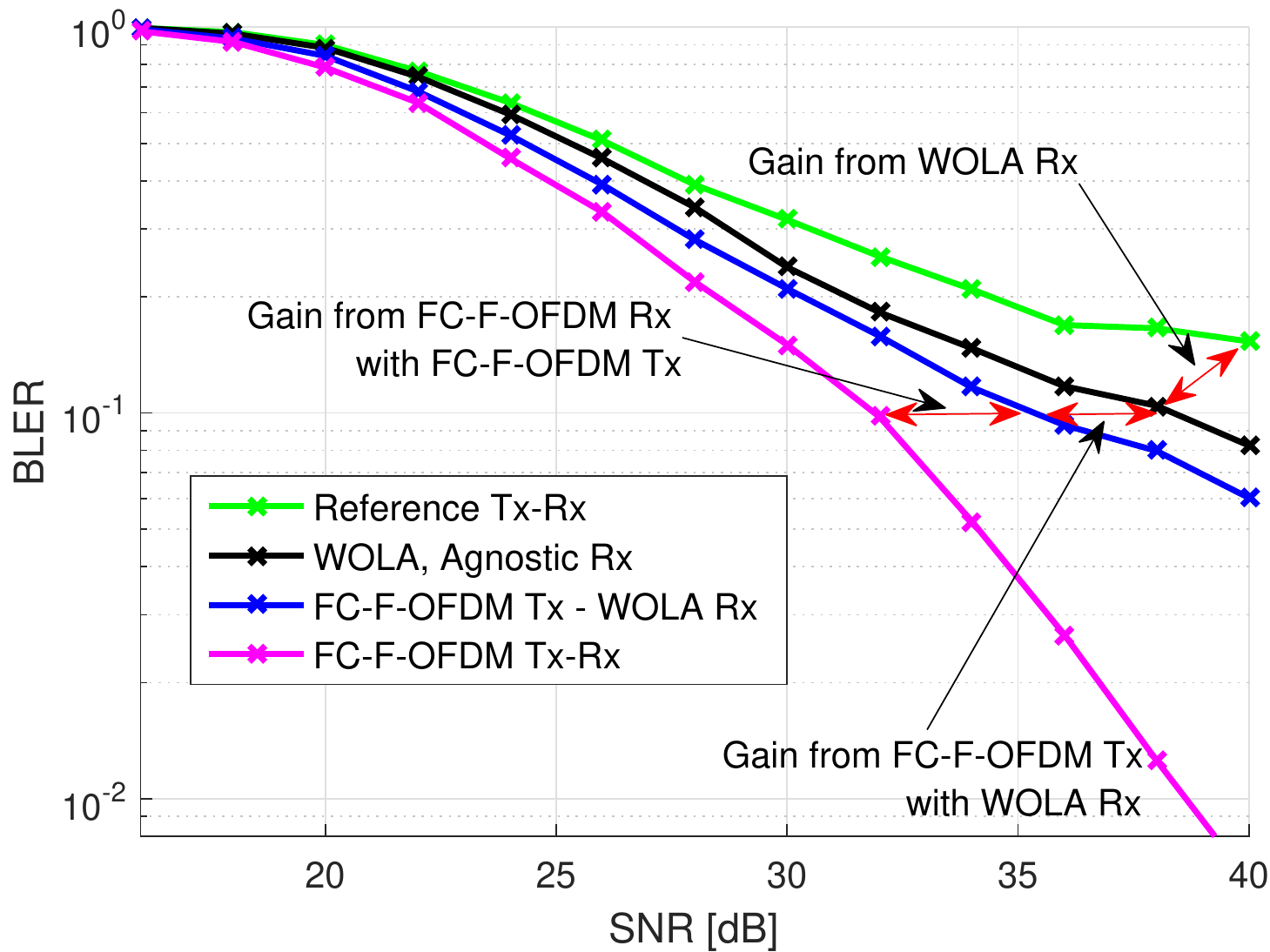}}
  \caption{In (a) the baseline DL radio link performance without interference is shown, and in (b)-(d) the DL radio link performance with 30~kHz subcarrier spacing based interfering signal (mixed numerology) and 180~kHz (1 PRB) guard band are given. The desired and interfering signals use 256-QAM and code rate R=4/5 in TDL-C~300~ns channel. In (b), the agnostic Tx and Rx approaches are compared, and in (c) the agnostic Rx performance is compared against matched Tx-Rx link. In (d), a predicted 5G NR radio link performance evolution is illustrated.}
  \label{fig:case2}
\end{figure*}

\subsubsection{Interference-free Baseline Performance}

\begin{figure*} 
  \centering
  \subfloat[][]{\includegraphics[angle=0,width=0.93\columnwidth]{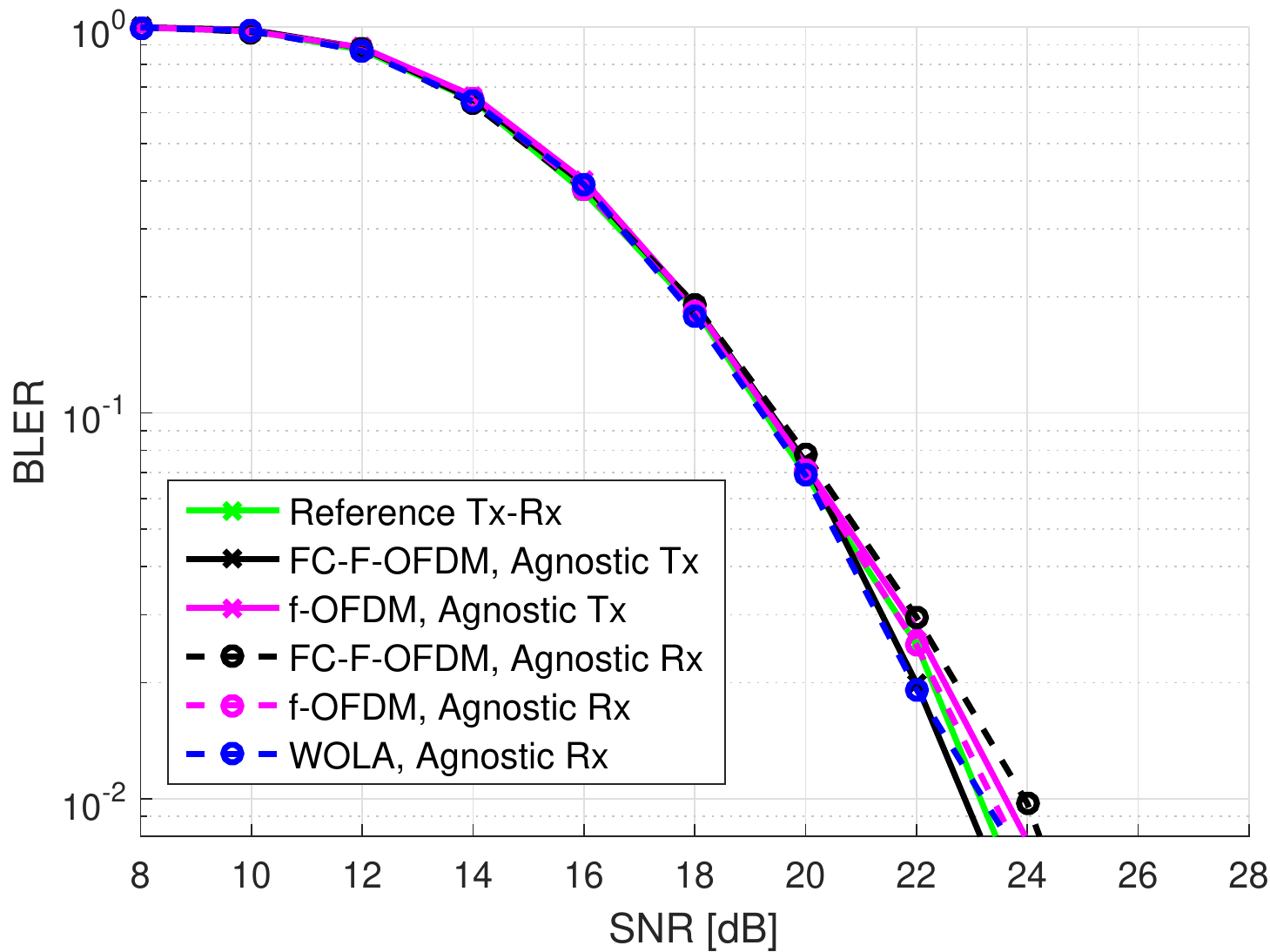}}
  \qquad
  \subfloat[][]{\includegraphics[angle=0,width=0.93\columnwidth]{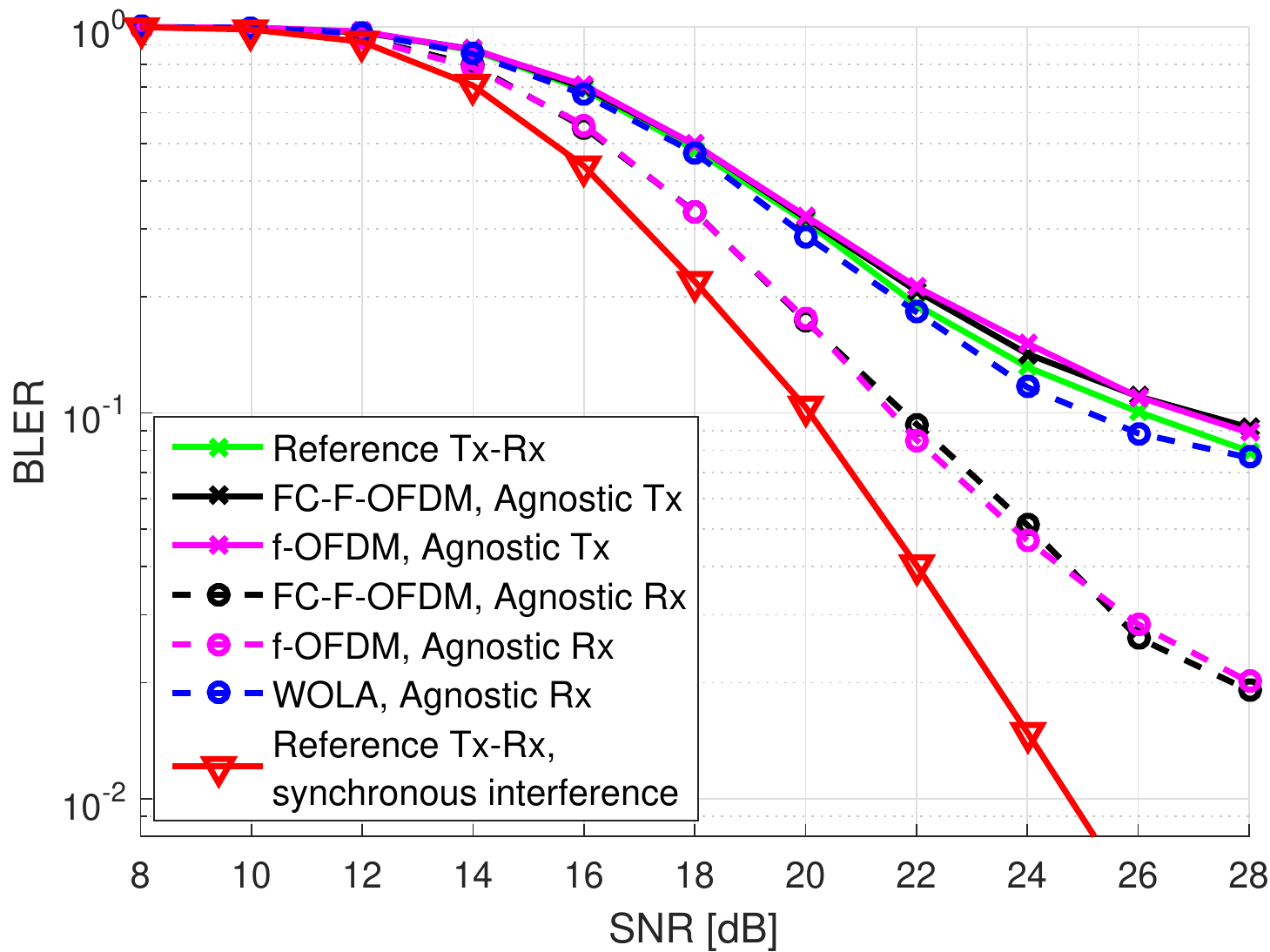}}
  \qquad
  \subfloat[][]{\includegraphics[angle=0,width=0.93\columnwidth]{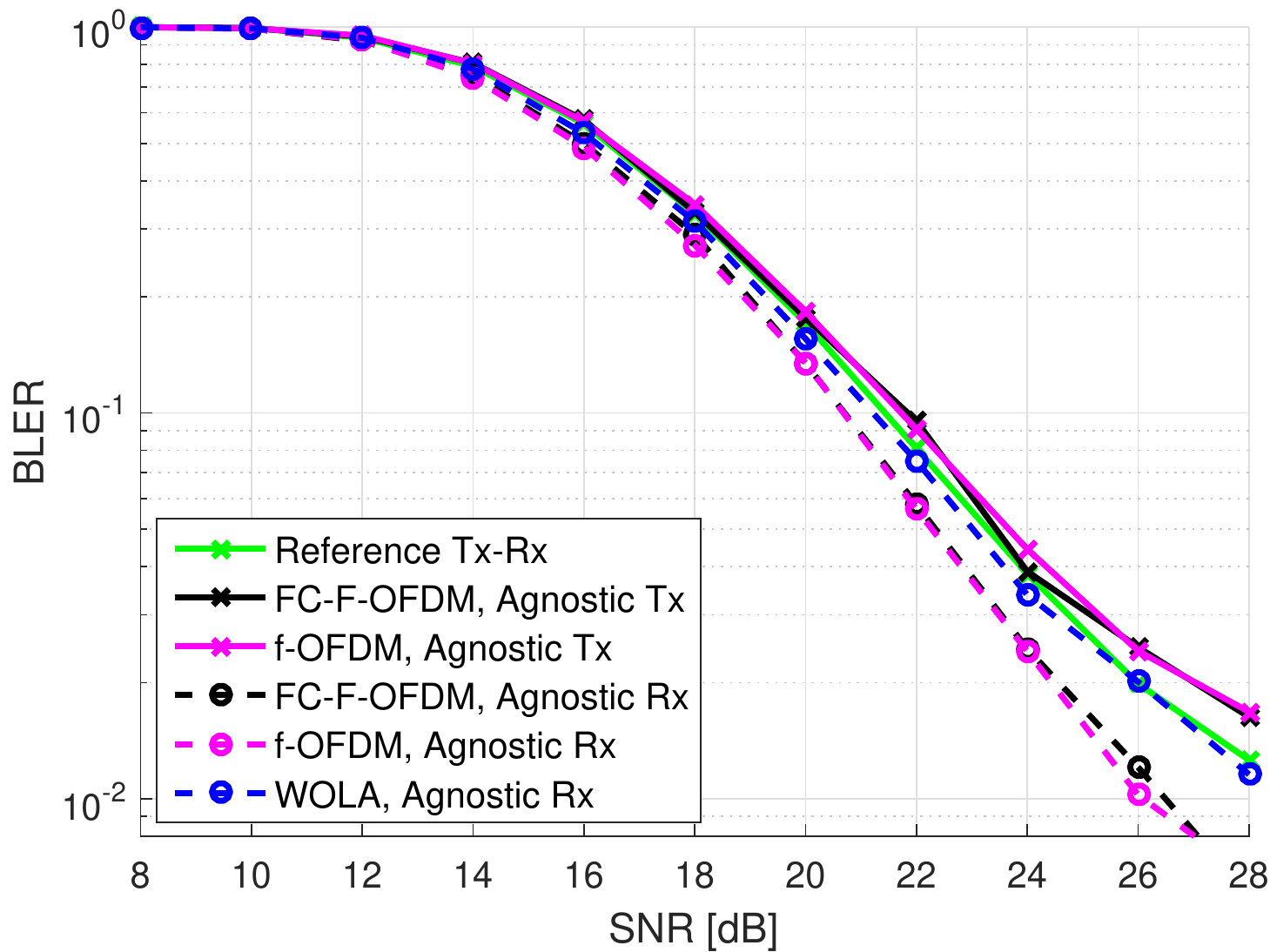}}
  \qquad
  \subfloat[][]{\includegraphics[angle=0,width=0.93\columnwidth]{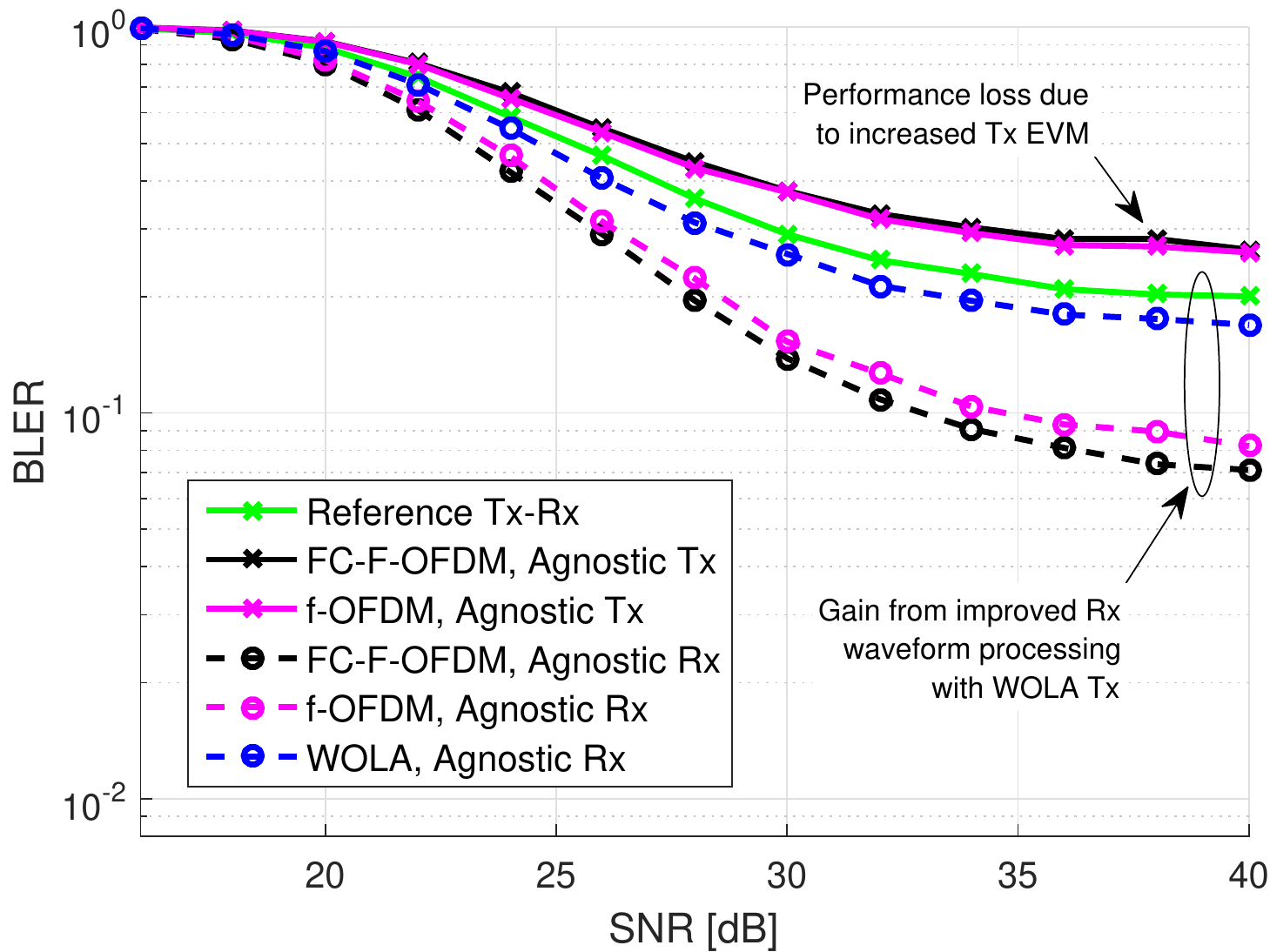}}
  \caption{Baseline radio link performance without interference in UL (a) with 64-QAM, code rate R=3/4 in TDL-C~1000ns channel, and in (b)-(c) the corresponding radio link performance assuming 30~kHz GB with (b) asynchronous interference and (c) mixed numerology interference. In (d), the UL mixed numerology performance with 256-QAM, R=4/5 modulation in TDL-C~300ns channel is shown assuming 30~kHz GB.}
  \label{fig:case3_4}
\end{figure*}

The first results correspond to interference-free DL and UL transmissions shown in Fig. \ref{fig:case2} (a) and Fig. \ref{fig:case3_4} (a), respectively. Any waveform processing aiming to improve performance in interference scenarios should not essentially degrade the performance in interference free DL case or synchronous UL case, which are currently the main use cases in eMBB services. In Fig. \ref{fig:case2} (a), it can be observed that without interference, the DL radio link performance is marginally degraded by the introduction of steep subband or BWP filtering. This is due to minor increase in inter-symbol-interference induced by filtering. In general, the new waveform processing techniques do not essentially limit the performance in interference-free scenarios. In the UL direction, as shown in Fig. \ref{fig:case3_4} (a), the differences are even smaller mainly due to the more robust 64-QAM modulation.

In the agnostic Tx cases, the assumption is that the Tx unit is using the waveform processing defined in the figure legend and the receiver is a reference Rx. In the agnostic Rx cases, in turn, the Rx unit implements the identified waveform processing technique while the transmitter is a reference Tx. For example, in Fig. \ref{fig:case2} (a), the "FC-F-OFDM, Agnostic Rx" case corresponds to FC-F-OFDM based Rx unit with reference waveform processing in the Tx unit (channel filtered CP-OFDM in DL), following the evaluation models and discussion given in Section \ref{sec:framework}. Notice also that the reference Tx-Rx and all agnostic Tx cases in Fig. \ref{fig:case2} (a) build on plain CP-OFDM Rx (the UE reference Rx), forming the basis for more evolved mixing of waveform processing solutions.

\subsubsection{DL Mixed Numerology Performance}

In Fig. \ref{fig:case2} (b)-(d), the DL radio link performance is given for a mixed numerology interference scenario, shown in Fig. \ref{fig:Tx_Rx_unit_testing} (c), where the desired 15 kHz subcarrier spacing based signal at BWP 1 is neighbored from right hand side by an interfering signal with 30 kHz subcarrier spacing at BWP 2. Both signals are assumed to contain 4 PRBs and use 256-QAM, R=4/5 modulation. In the DL mixed numerology case, when testing gNB Tx units (agnostic Tx cases), one can assume that all BWPs are processed with the selected waveform processing technique and therefore the desired and interfering signals build on the same waveform processing. When testing the Rx unit (agnostic Rx cases), the desired and interfering signals use the same reference Tx waveform processing and only the UE side Rx waveform processing is changed. In Fig. \ref{fig:case2}, the used GB corresponds to one 15~kHz subcarrier spacing based PRB.

In Fig. \ref{fig:case2} (b), the reference design does not achieve the 10\%~BLER target, and thus in practice a larger GB is required to support 256-QAM with code rate R=4/5 in DL mixed numerology scenario. Based on our results, the reference Tx-Rx achieves 10\%~BLER target with SNR of 38~dB or 36~dB with two or three PRB GB, respectively. Thus, for the reference Tx-Rx design and first generation 5G NR systems, larger GB than 1 PRB is required for sufficiently good link performance in mixed numerology DL scenario with high-order modulation and coding schemes. It is also observed that the improved spectral containment on the Tx side has relatively little effect, due to plain CP-OFDM receiver. On the other hand, when the UE Rx waveform processing is assumed to support highly selective filtering, a clear improvement in the performance is observed although the reference Tx unit does not improve the inband spectral containment of the transmitted signal.

When comparing matched link performance where similar waveform processing is assumed in Tx and Rx with agnostic Rx processing as shown in Fig. \ref{fig:case2} (c), we note that WOLA provides similar performance with matched link and agnostic Rx processing, while using a matched subband filtering scheme capable of providing a good spectral containment in both Tx and Rx improves the link performance compared to agnostic Rx performance. An important factor for clearly improved link performance in the filtering-based matched Tx-Rx cases is that the gNB Tx simultaneously improves the spectral containment of all BWPs, thus reducing the interference leakage between BWP~1 and BWP~2. From the evaluated waveforms, FC-F-OFDM is the only one achieving the interference free performance in DL matched Tx-Rx case. 

In general, it is most likely unrealistic to assume the first or even the second generation 5G NR UEs to support any DL Rx processing more complex than WOLA. On the other hand, the second generation gNB Tx could support FC-F-OFDM. Therefore, in Fig. \ref{fig:case2} (d), a possible evolution of 5G NR mixed numerology DL radio link performance is shown. Here reference Tx-Rx is shown together with reference Tx with WOLA Rx, FC-F-OFDM Tx with WOLA Rx, and FC-F-OFDM based Tx and Rx link. We observe that there is a clear gain in using WOLA in the Rx compared to reference Rx, and approximately 2~dB gain when combining FC-F-OFDM Tx with WOLA Rx. This gives a concrete vision and view how the 5G NR radio link performance can evolve over time by introducing gradually more and more complex, transparent physical layer processing in the gNB and UEs.

\subsubsection{UL Mixed-Numerology and Asynchronous Interference Performance}

In UL interference scenarios, illustrated in Fig. \ref{fig:Tx_Rx_unit_testing} (c), all signals are assumed to contain 4 PRBs and the interfering signals are always WOLA processed QPSK, R=3/4 modulated signals, following the reference Tx unit specifications defined in Section \ref{sec:framework}. The PA output power is always set such that 50\% of the EVM budget \cite{3GPPTS38101-1} is used by the polynomial PA model. By keeping the interfering signals fixed, we get a better understanding on the practical effects of certain waveform processing technique in the evaluated Tx or Rx unit. In practice, UEs from different vendors would anyway use different waveform processing techniques. 

\begin{table*}[t]
  \caption{Waveform processing complexities for different allocation sizes with different evaluated schemes.}
  \label{tab:results}
  \centering
  \footnotesize{
   \begin{tabular}{ccccccc}
      \toprule
      \multicolumn{1}{c}{} & 
      
      \multicolumn{2}{c}{FC-F-OFDM} &
      \multicolumn{2}{c}{WOLA} &
      \multicolumn{2}{c}{f-OFDM} \\  
      \cmidrule(lr){2-3}\cmidrule(lr){4-5}\cmidrule(lr){6-7}
      \multicolumn{1}{c}{Allocation} & 
      \multicolumn{2}{c}{Complexity} & 
      \multicolumn{2}{c}{Complexity} & 
      \multicolumn{2}{c}{Complexity} \\ 

      \multicolumn{1}{c}{size} & 
      \multicolumn{1}{c}{muls/act. SC} & 
      \multicolumn{1}{c}{relative to OFDM} &
      \multicolumn{1}{c}{muls/act. SC} & 
      \multicolumn{1}{c}{relative to OFDM} &
      \multicolumn{1}{c}{muls/act. SC} &
      \multicolumn{1}{c}{relative to OFDM} 
      \\
      \midrule
      1\,PRB & 1417.3 & $\times$2.4 &
      600.7 & $\approx \times$1.0 & 1139.0 & $\times$1.9 \\
      \midrule
      4\,PRBs & 354.3 & $\times$2.4 &
      150.2 & $\approx \times$1.0 & 284.8 & $\times$1.9\\
      \midrule
      50\,PRBs & 63.2 & $\times$5.3 & 12.0 & $\approx \times$1.0 & 954.5 & $\times$79.9 \\
      \midrule
      12$\times$4\,PRBs & 60.8 & $\times$4.9 & 12.5 & $\approx \times$1.0 & 284.8 & $\times$22.9 \\
      \midrule
      25 + 12\,PRBs & 58.4 & $\times$4.8 & 22.8 & $\times$1.9 & 638.4 & $\times$26.5 \\
      \bottomrule
    \end{tabular}}  
\end{table*}

In Fig. \ref{fig:case3_4} (b), the UL performance with asynchronous interference is shown. In this case the desired signal at subband 1 is neighbored on both sides by signals using the same numerology but which are shifted in time by 128~samples to model asynchronous operation. As in DL mixed numerology case shown in Fig. \ref{fig:case2}, the UE Tx waveform processing selection has relatively little effect on the performance whereas the gNB Rx waveform processing should build on highly selective subband filtering. In Fig. \ref{fig:case3_4}, we have included as a reference the performance with synchronous UL signals with the same GB. This provides a fair comparison point for the UL interference scenarios. We can observe, that with highly selective gNB Rx filtering the asynchronous UL performance is within 2~dB from the completely synchronous UL performance while the other asynchronous schemes are within 5~dB to 7~dB from the synchronous UL performance.

In Fig. \ref{fig:case3_4} (c), the UL performance with mixed numerology interference is shown. The desired signal with 15~kHz subcarrier spacing at BWP 1 is neighbored on both sides by interfering signals using a 30~kHz subcarrier spacing. The performance degradation is smaller than in asynchronous case with the same GB. This is mainly due to the lower power spectral density of the interfering signals due to wider bandwidth allocation. The same observations hold as in previous interference cases and we can see that the reference Tx-Rx setup is within 1.5~dB SNR gap from the reference single-numerology synchronous UL performance shown in Fig. \ref{fig:case3_4} (b). 

In Fig. \ref{fig:case3_4} (d), the UL mixed numerology performance evaluation is pushed further, with 30~kHz GB and 256-QAM, R=4/5. The observed performance degradation compared to 64-QAM in Fig. \ref{fig:case3_4} (c) is caused by the increased sensitivity of 256-QAM to inter-BWP interference. In the reference Tx-Rx and agnostic Tx cases, the receiver is always a channel filtered CP-OFDM receiver which does not attenuate the interference between BWPs, and thus the corresponding radio link performance saturates to a high BLER floor not reaching the 10\% target. Interestingly, in agnostic Tx cases, the BLER floor is even slightly higher compared to the reference Tx-Rx case, due to the Tx EVM increase caused by the selective BWP filtering as well as the fact that the interfering signals are fixed to WOLA processed CP-OFDM and hence do not change when target Tx unit is varied. Strictly-speaking, such observation can be made already in Fig. \ref{fig:case3_4} (b) and (c) but is further pronounced in (d) due to 256-QAM. Again, the agnostic Rx approach with highly selective Rx BWP filtering provides the best performance, with FC-F-OFDM being the best method similar to the DL studies. This is because the Rx BWP filtering efficiently suppresses the inter-BWP interference originating from BWPs 2 and 3.

The above finding that under reference processing methods in gNB Rx unit and in the Tx units of the baseline UE population, improving the Tx spectral containment of selected UE devices does not map to improved UL radio link performance is important and novel. Thus, the results indicate that special care in evaluating new Tx and Rx units in different scenarios with different reference Tx and Rx unit configurations is needed to understand the expected performance in practical deployments. At large, the obtained UL results indicate that in both mixed numerology and asynchronous interference scenarios considerable improvement in radio link performance can be achieved with \textit{agnostic} combination of unmatched Tx and Rx waveform processing solutions, with special emphasis on the gNB Rx unit side.

\subsection{Complexity Analysis and Comparison}

To understand the performance-complexity trade-offs related to the considered waveform processing techniques more thoroughly, we next address the involved computational complexities. Two basic complexity metrics are used, namely the number of real multiplications per active subcarrier and the relative complexity compared to plain CP-OFDM symbol processing with FFT size $N_\text{FFT}=1024$. The obtained complexity results are collected in Table \ref{tab:results}. The first four cases correspond to single numerology processing, with 15~kHz subcarrier spacing, where the $12 \times 4$ PRBs case corresponds to a concrete example of subband level filtering or windowing in the asynchronous UL access inside one BWP. The last line reflects an example mixed numerology case with two BWPs including $25 + 12$ PRBs with 15~kHz and 30~kHz subcarrier spacings, respectively. In all cases the shown complexities reflect a processing window of one 15~kHz CP-OFDM symbol.

In the single numerology scenarios, the WOLA scheme needs only a minor increase in complexity compared to the basic CP-OFDM, while f-OFDM and FC-F-OFDM are clearly more complex. Among the filtering schemes, f-OFDM implementation is slightly more effective in case of single or few narrow subbands, while for high numbers of subbands, or wider subbands, the FC-F-OFDM scheme is clearly more efficient in terms of the needed multiplication rates. 

In the mixed numerology scenario, WOLA has the largest relative increase in complexity compared to single numerology processing, as each numerology requires its own IFFT \cite{J:2013_Bala_WOLATxRx}. The f-OFDM, in turn, shows slightly increased complexity compared to the $12 \times 4$~PRBs case, but clearly lower complexity than in the 50~PRBs case modeling channel filtering like operation. Notably, FC-F-OFDM achieves even slighty smaller complexity than in the 50 and $12 \times 4$~PRBs cases.

The presented complexity results indicate that WOLA has the smallest complexity, however, as the number of simultaneously supported numerologies increases, the relative complexity difference between WOLA and different filtering-based approaches decreases. When considering also the presented radio link performance results, FC-F-OFDM is considered to provide the best complexity-performance tradeoff among the evaluated waveform processing solutions.

\section{Conclusion and Discussion}
\label{sec:conclusions}

In this article, 5G NR system evolution through transparent Tx and Rx processing enhancements was discussed and analyzed, with particular emphasis on new mixed numerology and asynchronous services. The baseline assumption is that CP-OFDM waveform is used, which can then be further filtered or windowed in the Tx or Rx unit at carrier, bandwidth part, or subband level. It was shown that different waveform signal processing techniques can be flexibly mixed, allowing to separately optimize complexity-performance trade-offs for transmitter and receiver implementations, and separate evolution paths for base-stations and user equipment. The observation that different Tx and Rx waveform processing algorithms can be flexibly mixed as long as they work efficiently also with plain CP-OFDM Rx or Tx, respectively, and that they provide similar performance with respect to matched Tx-Rx links is novel and generally less discussed in the available 5G NR literature. Furthermore, reference Tx and Rx waveform processing solutions for DL and UL, reflecting our view of the first phase 5G NR UE and gNB implementations, were defined and shown to provide realistic performance while ensuring also sufficient room for performance improvements with future device generations and categories in 5G NR system evolution. 

Based on the presented radio link performance results and complexity analysis, it is clear that highly-selective bandwidth part or subband filtering in the form of FC-F-OFDM should be applied in all Tx and Rx units in DL and UL, in the long run. The results also show that applying FC-F-OFDM in the Rx units has larger performance impact in both DL and UL, compared to Tx units. Due to the complexity increase compared to WOLA, FC-F-OFDM is most likely first implemented in the gNB side, where it can significantly improve the UL and DL radio link performance. Eventually, FC-F-OFDM should also be applied in the UE side to minimize inband interference and required GBs to maximize the 5G NR system throughput and spectral efficiency.

In general, the concept of transparent waveform processing is especially important in 5G NR mobile communications, where the flexible physical layer definition allows a single device to transmit or receive multiple bandwidth parts with different subcarrier spacings. Thanks to the transparent processing, each of these component signals may use different waveform processing technique on top of CP-OFDM to optimize the performance and complexity trade-offs per service. Furthermore, it may be envisioned that forthcoming 5G devices can adapt their waveform processing depending on the channel and interference conditions as long as the waveform processing itself is not specified, only the performance requirements, thus allowing to further improve the link performance in different scenarios. At large, the concepts put forward in this article provide a future-proof approach for 5G NR design, and simple and repeatable evaluation framework that can be used to assess new concepts for CP-OFDM based communications in 5G NR networks. 

\newpage

\bibliographystyle{IEEEbib}
\bibliography{main_arxiv}

\end{document}